\documentclass[journal,twocolumn,10pt,twoside]{IEEEtran}
\usepackage[T1]{fontenc}
\usepackage[utf8]{inputenc}
\usepackage{amsmath} 
\usepackage{amsfonts} 
\usepackage{amssymb} 
\usepackage{amsthm}
\usepackage{color}
\usepackage{url}
\usepackage{setspace} 
\usepackage{verbatim}
\usepackage{placeins}
\usepackage[linesnumbered,ruled]{algorithm2e}
\usepackage{multirow}
\usepackage{subcaption}
\usepackage{graphicx}
\usepackage{cite}
\usepackage{stackengine}

\renewcommand{\vec}[1]{{\bf{#1}}} 
 
\newcommand{\tran}{^{\mbox{\scriptsize T}}}
\newcommand{\herm}{^{\mbox{\scriptsize H}}}

\newcommand{\fro}[1]{\Vert #1\Vert_{\mbox{\scriptsize F}}}
\newcommand{\norm}[1]{\Vert #1\Vert}
\DeclareMathOperator*{\argmin}{arg\,min} 
\newcommand{\delequal}{\mathrel{\ensurestackMath{\stackon[1pt]{=}{\scriptstyle\Delta}}}}

\newcommand{\trace}[1]{\mathrm{tr}\left(#1\right)}

\newcommand{\phib}{\boldsymbol{\phi}}

\newcommand{\psib}{\boldsymbol{\psi}}

\def\green{\textcolor{green}}

\newcommand{\mybibliography}{\bibliography{jour_short,conf_short,References.bib}}

\title{Exploiting Spatial Correlation for Pilot Reuse in Single-Cell mMTC}

\author{Lucas~Ribeiro$^\dag$, Markus~Leinonen$^\dag$, Hanan~Al-Tous$^\star$, Olav~Tirkkonen$^\star$, and Markku~Juntti$^\dag$\\
        $^\dag$Centre for Wireless Communications, FI-90014, University of Oulu, Finland\\ 
        $^\star$Department of Communications and Networking, Aalto University, Finland\\
        e-mail: \{lucas.ribeiro,\,markus.leinonen,\,markku.juntti\}@oulu.fi,~\{hanan.al-tous,\,olav.tirkkonen\}@aalto.fi
        }

\begin{document}

\maketitle

\begin{abstract}
As a key enabler for massive machine-type communications (mMTC), spatial multiplexing relies on massive multiple-input multiple-output (mMIMO) technology to serve the massive number of user equipments (UEs). To exploit spatial multiplexing, accurate channel estimation through pilot signals is needed. In mMTC systems, it is impractical to allocate a unique orthogonal pilot sequence to each UE as it would require too long pilot sequences, degrading the spectral efficiency. This work addresses the design of channel features from correlated fading channels to assist the pilot assignment in multi-sector mMTC systems under pilot reuse of orthogonal sequences. In order to reduce pilot collisions and to enable pilot reuse, we propose to extract features from the channel covariance matrices that reflect the level of orthogonality between the UEs channels. Two features are investigated: \textit{covariance matrix distance} (CMD) feature and CMD-aided \textit{channel charting} (CC) feature. In terms of symbol error rate and achievable rate, the CC-based feature shows superior performance than the CMD-based feature and baseline pilot assignment algorithms.

\end{abstract}

\section{Introduction}

Massive machine-type  communications (mMTC) is a class of services planned for 5G and 6G systems to provide access for enormous numbers of connected devices, possibly on the order of tens of billions~\cite{Saad2020,Carvalho2017,Bockelmann2016}. Massive multiple-input multiple-output (mMIMO) technology has been proposed as a way to improve the spectral and energy efficiency of such systems by exploiting the spatial multiplexing provided by the large numbers of antennas~\cite{Larsson2014,Bjornson2016}. To take advantage of the boost in spectral and energy efficiency provided by large-scale antenna systems, channel state information (CSI) is required at the base station (BS). Uplink CSI is obtained through the transmission of pilot sequences from the user equipments (UEs) to the BS, which estimates the channel using the received signal by employing any estimation technique such as least squares or minimum mean square error. A time-division duplexing (TDD) protocol is likely to be implemented in mMIMO networks in order to minimize signaling overhead and to avoid downlink channel estimation by exploiting the reciprocity between the uplink and downlink channels~\cite{Al-hubaishi2019}.

One of the main challenges of mMTC is the design of access protocols that support the huge number of MTC devices. Due to the large number of UEs, resources have to be shared
, which precludes the allocation of orthogonal pilot sequences to all UEs. \textit{Pilot reuse} can be used to overcome the lack of resources, at the cost of leading to interference between UEs sharing the same pilot sequence, also known as \textit{pilot contamination}~\cite{Bjornson2017}.
While legacy systems rely on code and frequency multiplexing, mMTC can exploit spatial multiplexing through mMIMO to avoid/alleviate pilot contamination~\cite{Carvalho2017,Sanguinetti2020}. 

Different approaches to mitigate pilot contamination have been proposed in the literature. In~\cite{Akbar2018}, the authors proposed a location-aware pilot assignment for multi-cell systems with Rician fading channels which requires knowledge about the large-scale fading, angle of arrival (AoA), and the channel Rician factor. 
In~\cite{Al-hubaishi2019}, a pilot assignment scheme was devised to deal with pilot interference in multi-cell mMIMO based on the UEs' large-scale coefficients, which are assumed to be known in all cells. Their objective is to assign orthogonal pilot sequences to UEs with poor channel quality.
You {\it et al.}~\cite{You2015} proposed a pilot reuse strategy based on the channel covariance matrices for single-cell mMIMO systems with correlated channels. They proposed a pilot allocation algorithm that assigns orthogonal pilot sequences to UEs with similar channel covariance matrices, measured by the covariance matrix distance (CMD) metric.

Recently, we proposed a pilot reuse scheme in \cite{Ribeiro2020} that utilizes \textit{channel charting} (CC) to exploit the spatial information existing in CSI, aiming to maximize the AoA distances between the UEs using the same pilot sequence. CC is a framework proposed in~\cite{Studer2018} to estimate the relative position of devices in an unsupervised manner which maps the information obtained from the measured long-term CSI at the BS into a low-dimensional chart, in which the relative positions of UEs are preserved. 
Our results~\cite{Ribeiro2020} showed that CSI can be utilized to allocate the pilot sequences, and improve the channel estimation accuracy and symbol error rate (SER), for single-cell correlated channels systems.

In this paper, we propose pilot reuse strategies to cope with the pilot contamination in multi-sector single-cell mMIMO networks under spatially correlated channels. To exploit the degree of orthogonality between the UEs' channels, granted by the spatial correlation, the pilot reuse is performed based on two features encompassing the channel correlation, especially in the angular domain: 1) the CMD metric and 2) the CC created based on the CMD metric. Utilizing the proposed features along with the pilot allocation algorithm devised in~\cite{Ribeiro2020}, the numerical results show that the angular information can be exploited also for multi-sector systems with spatially correlated channels, resulting in enhanced achievable rate and SER compared to baselines.


\section{System model}

\begin{figure}[t]
    \centering
    \includegraphics[scale=0.3]{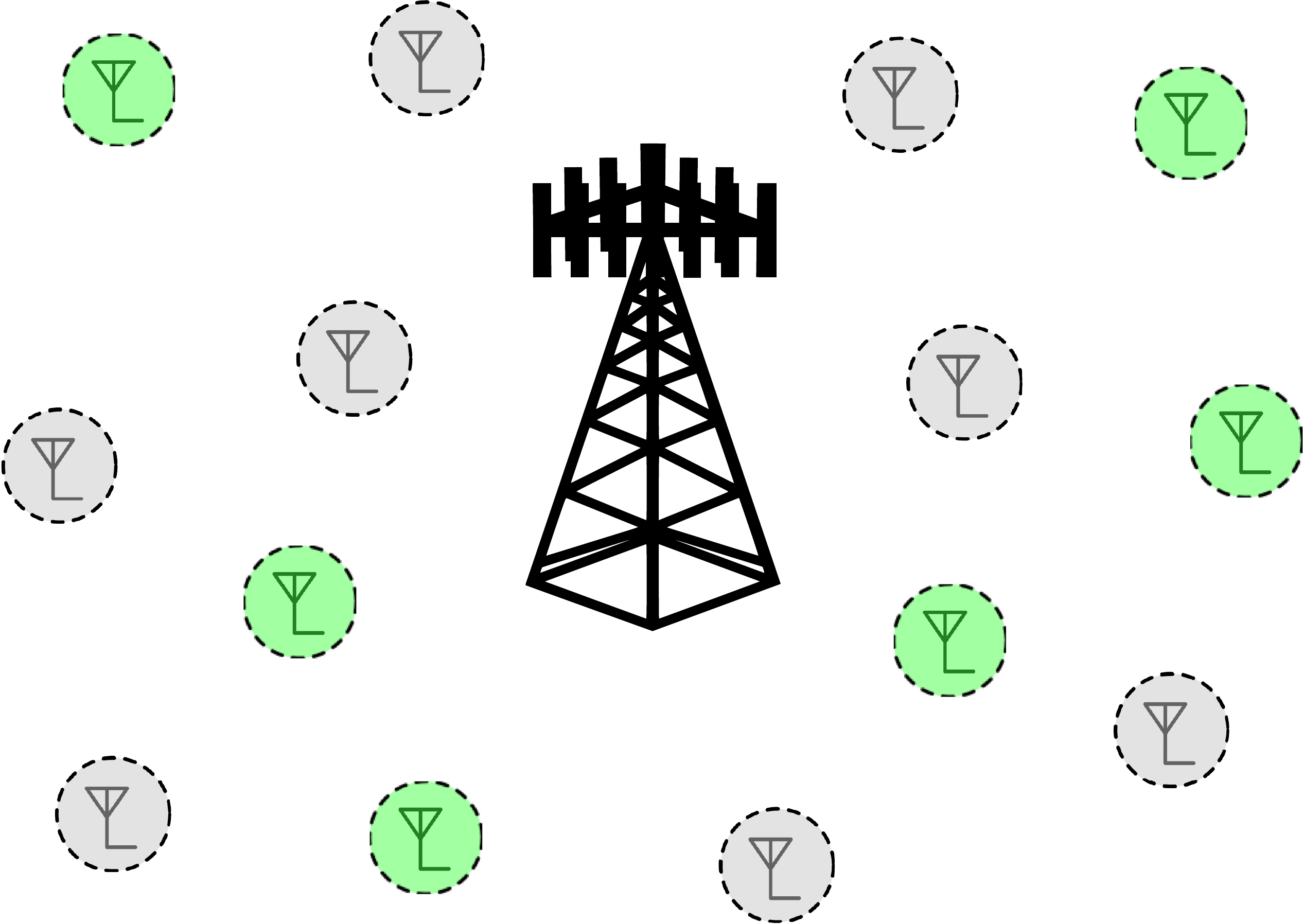}
    \caption{Uplink mMTC scenario with $K$ active (green) and $N-K$ inactive users (gray). The BS is equipped with $S=3$ $M$-element ULAs.}\vspace{-3mm}
    \label{fig:channel}
\end{figure}

We consider an uplink communication scheme with a set ${\mathcal{N}=\{1,\ldots,N\}}$ of $N$ single-antenna UEs uniformly distributed within a cell, from which only $K<N$ are randomly active at any given time. In order to focus on the development of pilot assignment strategies to improve the channel estimation accuracy, and, subsequently, enhance the system performance in mMTC, we assume that the BS knows the set of active UEs at each transmission instant\footnote{This assumption can be invoked by the fact that there exists a multitude of compressed sensing based approaches, e.g., \cite{chen2018sparse,Senel2018,Djelouat2020Joint}, to solve the UE activity detection problem. However, this is outside of the scope of this paper and is left for future work.}. 

The cell is divided into $S$ sectors\footnote{In practical scenarios, the coverage area is usually divided into three or six sectors of $120^\circ$ or $60^\circ$~\cite{Morales2019}. Therefore, we consider $S=3$ in the examples.} and the 
transmitted signal is received 
through $S$ uniform linear arrays (ULAs), each having $M$ antenna elements, as shown in Fig.~\ref{fig:channel}. The array response vector for a ULA is given by
\begin{equation} 
\vec{a}_{\mathrm{r}}(\theta) = \left[{1, e^{-j2\pi \Delta_{\mathrm{r}}\cos(\theta)}, \ldots, e^{-j2\pi (M-1)\Delta_{\mathrm{r}} \cos(\theta)}}\right]\tran 
\end{equation}
where $\Delta_{\mathrm{r}}$ is the normalized spacing between the antenna elements in units of wavelengths and $\theta$ is the AoA, i.e., the incident angle of the received signal on the antenna array\cite[Sec. 7.2.1]{TseBook}.

We adopt the one-ring channel model~\cite[p. 236]{Bjornson2017} which assumes that the multi-path components are concentrated around the UEs while the BS is located in an elevated position lacking scatterers close to it. Thus, the uplink channel vector for user $n\in\mathcal{N}$ associated with the ULA of sector $s\in \mathcal{S}=\{1,\ldots,S\}$ is modelled as a superposition of $L$ propagation paths as
\begin{equation}
\label{eq::hk-sector}
\vec{h}_{n,s} = \frac{1}{\sqrt{L}}\sum_{l=1}^{L}\sqrt{\beta_{n,s,l}}\alpha_{n,s,l}\vec{a}_{\mathrm{r}}(\theta_{n,s,l}),
\end{equation}
where $\alpha_{n,s,l}$ is the complex gain of the $l$th path assumed to be an independent and identically distributed (i.i.d.) complex Gaussian random variable with zero mean and $\mathbb{E}\{|\alpha_{n,s,l}|^2\}=1$. In \eqref{eq::hk-sector}, the large-scale propagation effects and the BS antenna gain in the channel are captured in $\beta_{n,s,l} \in \mathbb{R}$, which follows the free space path loss model as described in~\cite[Eq.~(2.7)]{Goldsmith2005}, which is defined as
 \begin{equation}
 \beta_{n,s,l} \delequal 10^{\frac{G_{\mathrm{A}}(\theta_{n,s,l})}{10}}\left(\frac{\lambda}{4\pi d_{n}}\right)^2, 
 \label{eq::beta}
 \end{equation}
where $\lambda$ is the wavelength and $d_{n}$ is the distance between UE $n$ and the BS. In~\eqref{eq::beta}, $G_{\mathrm{A}}(\theta_{n,s,l})\in \mathbb{R}$ represents the antenna gain through the $l$th path for user $n$ at the ULA $s$, and is given in dB by~\cite[Table 7.1-1]{TR36873,Morales2019} 
\begin{equation}
    G_{\mathrm{A}}(\theta_{n,s,l}) = G_{\mathrm{A_{max}}} {-\textrm{min}\left[12\left(\frac{\theta_{n,s,l}}{\theta_{3\textrm{dB}}}\right)^2,\;A_\textrm{max}\right]},
    \label{eq::AntennaGain}
\end{equation} 
where $G_{\mathrm{A_{max}}}$ is the maximum antenna gain, $\theta_{3\textrm{dB}}$ is the half power beamwidth, $A_\textrm{max}$ is the maximum attenuation in dB at the ULAs, and $\theta_{n,s,l}$ is the AoA of the $l$th path which is modelled as an i.i.d. random variable with uniform distribution $U(\theta_{n,s}^{\mathrm{min}},\theta_{n,s}^{\mathrm{max}})$, with $\theta_{n,s}^{\mathrm{min}}=\Bar{\theta}_{n,s}-\sqrt{3}\sigma_{\theta}$ and $\theta_{n,s}^{\mathrm{max}}=\Bar{\theta}_{n,s}+\sqrt{3}\sigma_{\theta}$. Here, $\Bar{\theta}_{n,s}\in [0,2\pi]$ is the incident angle between user $n$ and the ULA of sector $s$, $\sigma_{\theta}$ is the angular standard deviation, which specifies the AoA interval $\mathcal{A}_{n,s}=[\theta_{n,s}^{\mathrm{min}},\theta_{n,s}^{\mathrm{max}}]$ for the incoming multi-path components arriving from user $n$ at the ULA of sector $s$.

\section{Pilot Transmission and Channel Estimation}
\begin{figure}
    \centering
    \includegraphics[scale=0.45]{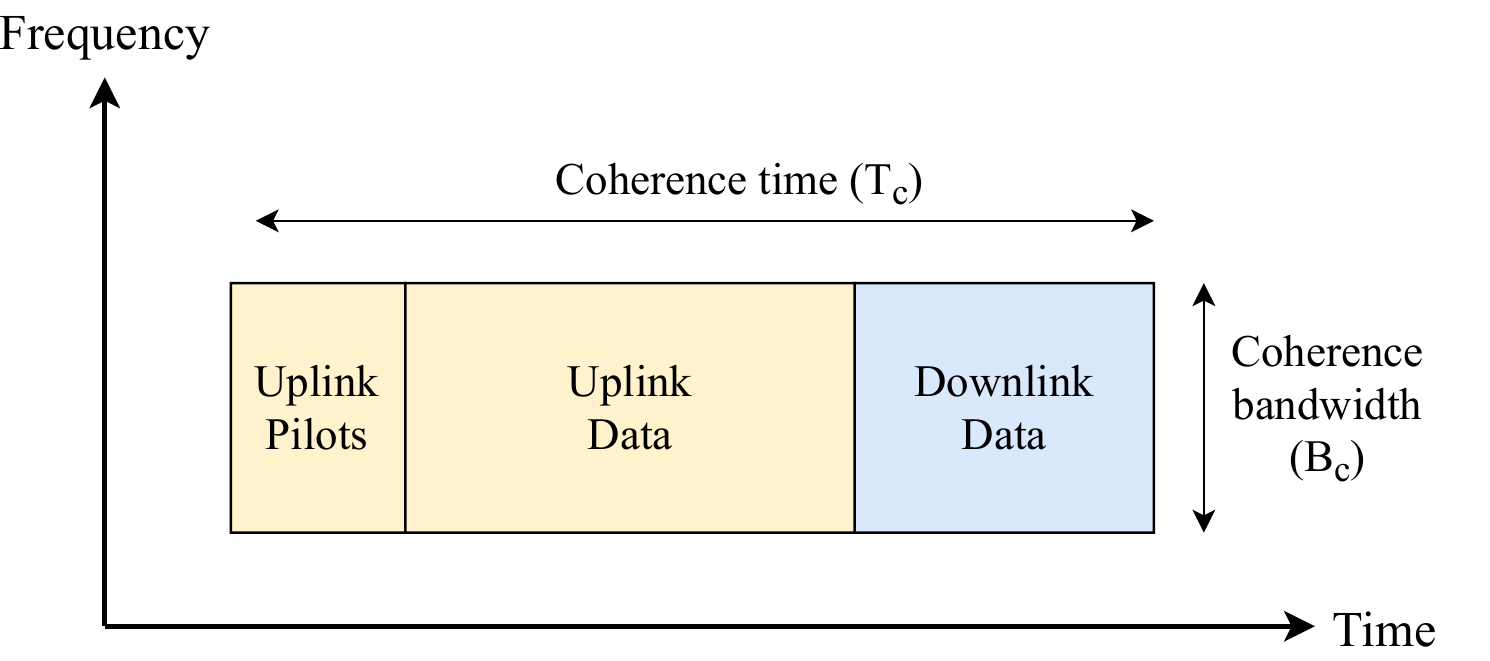}
    \caption{The channel response is time-invariant and frequency-flat within each coherence block. At the beginning of each coherence block, $\tau$ symbols are sent to estimate the channel.}\vspace{-2mm}
    \label{fig:TDD}
\end{figure}

To exploit the symmetry between downlink and uplink channels, we consider the TDD protocol, as depicted in Fig.~\ref{fig:TDD}. We assume that during one coherence block, the channels are time-invariant and flat-fading. At each coherence block, the set of active users transmit $\tau$ known symbols, of equal power $p_{\mathrm{u}}$, to the BS for channel estimation. Right after transmitting the pilot symbols, the $K$ active UEs transmit their data to the BS. Then, the remaining time, within the coherence block, is used for downlink communication.


We consider that the UEs are assigned with pilot sequences taken from a pool of $\tau$ \textit{orthogonal} sequences. However, due to a vast number of UEs in mMIMO networks, we consider that $\tau\ll N$, i.e., global \textit{pilot reuse} is employed in the cell area. Consequently, in the channel estimation phase, the same pilot is shared by $K/\tau$ UEs on the average. Let $\mathcal{K}=\{1,\ldots,K\}\subseteq\mathcal{N}$ represent the set of active UEs at a given transmission interval. Let ${\mathcal{T}=\{1,\ldots,\tau\}}$ be the set of indices of available pilot sequences. User $k\in\mathcal{K}$ transmits a pilot signal ${\vec{\psib}_k=\sqrt{p_{\mathrm{u}}}\phib_{\pi_k}}$,
where $\pi_k\in\mathcal{T}$ is the index of the pilot sequence assigned to UE $k$ and $\phib_{\pi_k}\,\in\,\mathbb{C}^{\tau}$ is the corresponding pilot sequence from the orthogonal pilot book $\vec{\Phi}=[\vec{\phib}_1,\ldots,\vec{\phib}_{\tau}] \in \mathbb{C}^{\tau\times \tau}$. We define the set of UEs sharing the same pilot sequence as UE $k$, including UE $k$ itself, as ${\mathcal{G}_k=\{j\mid j\in\mathcal{N}, \, \pi_j=\pi_k\}}$.

The received signal for channel estimation at the ULA of sector $s$, $\vec{Y}_{s}=\left[\vec{y}_{s}^1,\ldots,\vec{y}_{s}^\tau\right]\in\mathbb{C}^{M \times \tau}$, can be written as 
\begin{equation}
\vec{Y}_{s} = \vec{H}_{s}\vec{\Psi} + \vec{N}_{s},
\label{eq::receivedSignal}
\end{equation}
where $\vec{H}_{s}=\left[\vec{h}_{1,s},\ldots,\vec{h}_{K,s}\right]\in\mathbb{C}^{M \times K}$ is the channel matrix for the active UEs, $\vec{\Psi}=\left[\vec{\psib}_1,\ldots,\vec{\psib}_K\right]\tran\in\mathbb{C}^{K\times \tau}$ is the pilot signal matrix, and $\vec{N}_{s}=\left[\vec{n}_{s}^{1},\ldots,\vec{n}_{s}^\tau\right]\in\mathbb{C}^{M \times \tau}$ is the noise matrix. We model the noise as an i.i.d. complex Gaussian random variable ${\vec{n}_{s} \sim \mathcal{CN}(0,\sigma_\mathrm{n}^2)}$, where $\sigma_\mathrm{n}^2$ is the noise power at each antenna element.

We consider that the linear minimum mean square error (LMMSE) receiver is deployed at the BS to jointly estimate the active UEs' channel vectors across the $S$ sectors. Therefore, the compound channel $\vec{h}_{k}\in\mathbb{C}^{MS}$
between UE $k\in\mathcal{K}$ and the BS is given by
 \begin{equation}
\label{eq::hk}
\vec{h}_{k} =
\begin{bmatrix}
\vec{h}_{k,1} \\
\vdots\\
\vec{h}_{k,S}
\end{bmatrix},
\end{equation}
with covariance matrix $\vec{R}_{k}=\mathbb{E}[\vec{h}_{k}\vec{h}_{k}\herm]\in\mathbb{C}^{MS\times MS}$. The received pilot signal, 
$
 \vec{Y}=\left[\vec{Y}_1\tran,\ldots,\vec{Y}_S\tran\right]\tran\in\mathbb{C}^{MS\times\tau}.
$

We assume that the channel covariance matrices of all UEs (active and inactive ones) are known at the BS. In order to be able to 
retrieve the UEs' individual signals from the received compound signal, the BS needs to keep track of $\vec{R}_n,\,\forall n \in \mathcal{N}$. In practice, an initial training phase is required to obtain the first estimate for the covariance matrices. After this initial phase, the BS can keep updating each $\vec{R}_{n}$ based on the estimated channel.

The LMMSE estimate of the communications channel between user $k$ and the BS, $\vec{h}_{k}$ in \eqref{eq::hk}, is given as~\cite{You2015},
\begin{equation}
\label{eq::hk_hat}
    \vec{\hat{h}}_k = \vec{R}_{k}\vec{Q}_{k}^{-1}\vec{y}_{k}^{\mathrm{p}}.
\end{equation}
Here, $\vec{y}_{k}^{\mathrm{p}}$ represents the processed received signal for UE $k$ after correlating the received signal with the pilot sequence assigned to user $k$, i.e.,
\begin{equation}
\begin{split}
    \vec{y}_{k}^{\mathrm{p}} &= \frac{1}{p_{\mathrm{u}} \tau}\vec{Y}\vec{\psib}_{k}^{*} \\
    &=\vec{h}_k+\underbrace{\sum_{j\,\in\, \mathcal{I}_k}{\vec{h}_j}}_\text{Pilot Interference}+\frac{1}{p_{\mathrm{u}}\tau}\vec{N}\vec{\psib}_{k}^{*} ,
\end{split}
\label{eq::LS}
\end{equation}
where $\mathcal{I}_k=\{j|j\neq k,\,j\in\mathcal{G}_k\cap\mathcal{K}\}$ is the set of interfering users to user $k$,
and $\vec{Q}_{k}\in\mathbb{C}^{MS\times MS}$ in~\eqref{eq::hk_hat} is the covariance matrix of the received signal, given as
\begin{equation}
    \vec{Q}_{k}=\vec{R}_k+\sum_{j\in \mathcal{I}_k}\vec{R}_j + \frac{\sigma_\mathrm{n}^2}{p_u\tau}\vec{I}_{MS}.
    \label{eq::Rerror}
\end{equation}

Due to the orthogonality principle of the MMSE estimator~\cite[Sec. 12.4]{Kay1993}, the channel estimation error, $\vec{\tilde{h}}_{k}\sim \mathcal{CN}(\vec{0},\vec{R}_{\tilde{\vec{h}}_{k}})$, is independent of $\vec{\hat{h}}_{k}$. Therefore, we can decompose the channel $\vec{h}_{k}$ as $\vec{h}_{k}=\vec{\hat{h}}_{k}+\vec{\tilde{h}}_{k}$. Thus, the error covariance matrix for user $k$ is~\cite{Studer2018} 
\begin{equation}
    \vec{R}_{\tilde{\vec{h}}_{k}}=\vec{R}_{k}-\vec{R}_{k}\vec{Q}_{k}^{-1}\vec{R}_{k}.
    \label{eq::err-covariance}
\end{equation}

\section{Uplink Data Transmission}

Let $\vec{x}=\left[x_1,\ldots,x_K\right]\tran\in\mathbb{C}^{K}$ be the transmitted symbol vector at a given time instant. The corresponding received signal at the BS, $\vec{y}_{}=[\vec{y}_{1}\tran,\ldots,\vec{y}_{s}\tran]\tran \in \mathbb{C}^{MS}$, is given by
\begin{equation}
\vec{y}_{} = \vec{H}_{}\vec{x} + \vec{n}_{},
\end{equation}
where $\vec{H}_{}=\left[\vec{h}_{1},\ldots,\vec{h}_{K}\right]\in\mathbb{C}^{MS\times K}$ is the channel matrix, and $\vec{n}_{}=[\vec{n}_{1}\tran,\ldots,\vec{n}_{s}\tran]\tran \in \mathbb{C}^{MS}$ is the noise vector at the $M\times S$ receiver antenna elements. We assume that all UEs transmit with the same power, i.e., $p_{\mathrm{u}}=|x_k|^2$ is the transmit symbol power for user $k$.

Given the estimated channel $\vec{\hat{H}}=[\vec{\hat{h}}_{1},\ldots,\vec{\hat{h}}_{K}]\in\mathbb{C}^{MS\times K}$, we use the LMMSE receiver, $\vec{w}_k\in\mathbb{C}^{MS}$, derived in~\cite{You2015},
\begin{equation}
    \vec{w}_k=\left(\vec{\hat{H}}\vec{\hat{H}}\herm+\sum_{k=1}^{K}{\vec{R}_{\tilde{\vec{h}}_k}}+\frac{\sigma_\mathrm{n}^2}{p_u}\vec{I}_{MS}\right)^{-1}\vec{\hat{h}_k},
    \label{eq::LMMSECombinervector}
\end{equation}
which takes into account the error covariance matrix $\vec{R}_{\tilde{\vec{h}}_k}$.

Therefore, the received symbol vector $\hat{\vec{r}}\in\mathbb{C}^{MS}$, after employing the LMMSE receiver $\vec{W}\in\mathbb{C}^{MS\times K}$, is given by
\begin{equation}
    \hat{\vec{r}}=\vec{W}\herm \vec{y}.
\end{equation} 

Given the communication model depicted in Fig.~\ref{fig:TDD}, the uplink spectral efficiency is lower bounded by~\cite[Th. 4.1]{Bjornson2017}
\begin{equation}\label{eq::net-se}
    R^{\mathrm{up}}_k=\left(1-\frac{\tau}{T_c}\right)R^{\mathrm{ach,up}}_k,
\end{equation}
where $T_c$ is the coherence time and 
the corresponding achievable uplink rate for user $k$, $R^{\mathrm{ach,up}}_k$, is expressed as
\begin{equation}
    R^{\mathrm{ach,up}}_k=\mathbb{E}\left\{\log_2{\left(1+\gamma_k^{\textrm{up}}\right)}\right\}, 
\label{eq::Rachievable}
\end{equation}
where the expectation is taken over the channel realizations,
and $\gamma_k^{\textrm{up}}$ is the \textit{instantaneous} uplink signal-to-interference-and-noise ratio (SINR), given as
\begin{equation}
    \gamma_k^{\textrm{up}}=\frac{|\vec{w}_k\herm\vec{\hat{h}}_k|^2}{\vec{w}_k\herm\left(\sum_{j\neq k}{\vec{\hat{h}}_j\vec{\hat{h}}_j\herm+\sum_{n=1}^{K}{\vec{R}_{\vec{\tilde{h}}_n}}+\frac{\sigma_\mathrm{n}^2}{p_u}\vec{I}_{MS}}\right)\vec{w}_k}.
    \label{eq::SINR_instantaneous}
\end{equation}

\section{Pilot Reuse Algorithm}
The coordination of pilot assignment is crucial for mMTC. Due to the massive amount of UEs in such systems, the reuse of the orthogonal pilot sequences in the same cell becomes inevitable. As highlighted by Björnson in~\cite[p. 246]{Bjornson2017}, the strongest interference usually originates from UEs in the same cell. Therefore, a proper assignment of the pilots is essential to mitigate the pilot contamination.

Next, we tackle the pilot contamination problem by proposing pilot reuse strategies that exploit the spatial information present in CSI through the second-order statistics of the radio environment. In particular, we propose two features encompassing the channel correlation, especially in the angular domain, for the pilot reuse: 1) the CMD metric and 2) the CC created based on the CMD metric. These features are fed to a modified version of our greedy pilot allocation algorithm proposed in~\cite{Ribeiro2020} (Algorithm~\ref{alg:NNPA} detailed in Sect.\ \ref{subsec:NNPA}), which completes the pilot assignment, aiming at minimizing the pilot interference between the users sharing the same pilot. 



The goal is to assign orthogonal pilot sequences to users with overlapping AoA intervals to avoid pilot contamination. To capture the spatial information embedded in the second-order statistics and reveal the angular domain relationship among the UEs, we use the CMD~\cite{Herdin2005} metric. Thus, we compute a dissimilarity matrix $\vec{D}\,\in\mathbb{R}^{N\times N}$ and define the dissimilarity measure associated with UE $n$ and $j$ as
\begin{equation}\label{eq::CMD}
    d_{n,j}=1-\frac{\trace{\vec{R}_n\herm\vec{R}_j}}{\fro{\vec{R}_n}\fro{\vec{R}_j}}.
\end{equation}

Next, we elaborate on two different features that utilize the CMD metric in \eqref{eq::CMD} as a basis. The features will serve as input to our \textit{Nearest Neighbor Pilot Assignment} Algorithm 1.

\subsection{CMD-aided Pilot Assignment}
Since the dissimilarity matrix $\vec{D} =[\vec{d}_1,\ldots,\vec{d}_N]$ carries the spatial information needed to allocate the pilot sequences, we propose to directly utilize the distance vectors, $\vec{d}_n = [d_{n,1},\ldots,d_{n,N}]\tran\,\in\mathbb{R}^{N}$, as input features, ${\vec{f}_n}\,\in\mathbb{R}^{N}$, to the proposed pilot allocation algorithm as detailed in~\ref{subsec:NNPA}. Thus, we form $\vec{f}_n$ by setting 
$\vec{f}_n = \vec{d}_n$. After getting $\vec{f}_n$ we apply the Algorithm~\ref{alg:NNPA} to allocate the pilots.

\subsection{CC-aided Pilot Assignment}

CC is as a framework developed in~\cite{Studer2018} to generate unsupervised radio environment mappings utilizing the CSI. The idea behind CC is to find a suitable feature from CSI and then apply a dimensionality reduction (DR) technique to get a lower dimensional embedding, which preserves the relative position of the UEs.
Given the CMD distance in~\eqref{eq::CMD}, we propose to utilize it as an input to CC framework. 
The target is to obtain a feature that well characterizes the angular domain structures in the multi-sector multi-user scenario, in order to feed the pilot allocation Algorithm~\ref{alg:NNPA}.

As shown in~\cite{Ribeiro2020}, for single-cell scenarios, CC can be used to coordinate the pilot assignment under pilot reuse regimes, to achieve better network performance than other methods from the literature. In~\cite{Ribeiro2020}, we have utilized CC to retrieve the angular relationship between the UEs by applying the discrete Fourier transform on the channel covariance matrices and then taking the absolute value.
Here, we differently propose to use the CMD dissimilarity matrix  
$\vec{D}$ in~\eqref{eq::CMD} as an input to the DR technique, to construct the CC. Thus, we apply a function $\mathcal{C}$ that generates CC by mapping these high-dimensional features $\vec{d}_n\in\mathbb{R}^{N}$ to a low-dimensional domain, i.e.,
\begin{equation}
  \mathcal{C}:\vec{d}_n\mapsto\vec{f}_n,\quad n\in\mathcal{N},  
  \label{eq::cc-mapping}
\end{equation}
where ${\vec{f}_n\in\mathbb{R}^{C}}$ is the point in the $C$-dimensional
CC corresponding to $\vec{d}_n$, where typically $C=2$ or $C=3$.

Several unsupervised DR techniques have been proposed to map the extracted features into a lower dimension embedding\cite{Studer2018,Lei2019,Huang2019,Agostini2020}. We propose the use of Laplacian Eigenmaps (LE) as a DR technique as it aims to preserve the local structure of the high-dimensional embedding by minimizing the distances between data points and its $\nu$ nearest neighbors, where $\nu$ is a design parameter. Note that the CC method is not limited to utilize LE, and different DR techniques could be used as well. Clearly, as the low-dimensional embedding is not unique, they may lead to different pilot assignments, thus different network performance.

\subsection{Nearest Neighbor Pilot Assignment Algorithm}
\label{subsec:NNPA}

\begin{algorithm}[t]
\caption{Nearest neighbor pilot assignment}
\label{alg:NNPA}
\SetAlgoNoLine
\DontPrintSemicolon
\SetKwInOut{Input}{Input}\SetKwInOut{Output}{Output}
\SetKwFor{For}{for}{do}{end for}

\Input{1) The set of UEs $\mathcal{N}=\{1,\ldots,N\}$, 2) the index set of orthogonal pilots $\mathcal{T}=\{1,\ldots,\tau\}$, and 3) the suitable features $\vec{f}_n,\, n \in\,\mathcal{N}$.}

\Output{A pilot assignment $\vec{\Psi}=\left[\vec{\psib}_1,\ldots,\vec{\psib}_K\right]\tran$.}

Initialize the set of unassigned UEs $\mathcal{N}^{\mathrm{un}}=\mathcal{N}$, and the set of unassigned pilots $\mathcal{T}^{\mathrm{un}}=\mathcal{T}$.\;

Select a random UE $n$ and initialize the auxiliary variable $n'$ with it, i.e., $n'=n$.\;

Assign $\phib_1$ to user $n$ and update the set of unassigned UEs and pilots, i.e., $\mathcal{N}^{\mathrm{un}}\gets \mathcal{N}^{\mathrm{un}}\setminus \{n\}$ and $\mathcal{T}^{\mathrm{un}}\gets \mathcal{T}^{\mathrm{un}}\setminus \{1\}$.\;

Initialize the auxiliary variable: $p=2$.\;
\While{$\mathcal{N}^{\mathrm{un}}\neq \emptyset$}{
  \If{$\mathcal{T}^{\mathrm{un}}=\emptyset$}{
  Reinitialize: $\mathcal{T}^{\mathrm{un}}=\mathcal{T}$ and $p=1$.\;
   }
   
  Assign pilot $\phib_{p}$ to user $n$, i.e., $\vec{\psib}_k=\phib_p$, that satisfies $n=\underset{n\in \mathcal{N}^{\mathrm{un}}}{\argmin}\Vert \vec{f}_n-\vec{f}_{n'}\Vert^2$.\;
  
  Update the set of unassigned UEs, $\mathcal{N}^{\mathrm{un}}\gets \mathcal{N}^{\mathrm{un}}\setminus \{n\}$, and the set of unassigned pilots, $\mathcal{T}^{\mathrm{un}}\gets \mathcal{T}^{\mathrm{un}}\setminus \{p\}$.\;
  
  Update $n'=n$ and $p=p+1$.\;
 }
\end{algorithm}

After extracting the CMD and CMD + CC features, we apply a slightly modified version of the low-complexity pilot algorithm proposed in~\cite{Ribeiro2020} to account for the different input features and incorporate the inactive users as well, as shown in Algorithm~\ref{alg:NNPA}.  
The main premise of the algorithm is that the features preserve the angular relationship among UEs. Thus, the greedy algorithm allocates the orthogonal pilot sequences in an \textit{ordered} way $\phib_1,\phib_2,\ldots,\phib_\tau,\phib_1,\phib_2,\ldots$ by finding the unassigned UE with the small Euclidean distance between the features, aiming at maximizing 
the distances between the same pilot sequences.

The first step of Algorithm~\ref{alg:NNPA} is to allocate the first pilot sequence, $\phib_1$, to one of the users. Then, it greedily allocates the next orthogonal pilot sequence, $\phib_2$, to the unassigned UE with the smallest distance to the previously allocated user $n'$, i.e., it finds $n$ that minimizes $\norm{\vec{f}_n-\vec{f}_{n'}}^2$. Then, it repeats this process until all orthogonal pilot sequences have been allocated. Once the last orthogonal pilot sequence $\phib_\tau$ has been allocated, after $\tau$ iterations, it starts to reuse pilot sequences by allocating the first sequence $\phib_1$ to the closest unassigned UE from the precedent allocated user. It repeats this process until all UEs have been assigned pilot sequences.

\section{Numerical and Simulation results}
\label{sec:results}

We consider $N=512$ UEs uniformly distributed in a cell with $K=64$ active UEs at any given time, and a BS equipped with $S=3$ $M$-element ULAs, with $M=64$ critically spaced, $\Delta_r = 0.5$, elements. The propagation channel between each user and the BS consists of $L = 200$ paths with a fixed angular standard deviation $\sigma_\theta=15^{\circ}$. For the antenna parameters in~\eqref{eq::AntennaGain}, we set the maximum antenna gain as $G_{\mathrm{A_{max}}}=0$~dB, the maximum attenuation as $A_{\mathrm{max}}=30$~dB, and the half-power beamwidth as $\theta_{\mathrm{3dB}}=65^\circ$~\cite{TR36873}. We use binary phase shift keying (BPSK) for the channel estimation and quadrature phase shift keying (QPSK) for the data transmission.

\begin{figure}[t]
\centering
\begin{subfigure}{0.45\textwidth}
    \includegraphics[width=\textwidth,trim=1mm 0mm 10mm 2mm,clip]{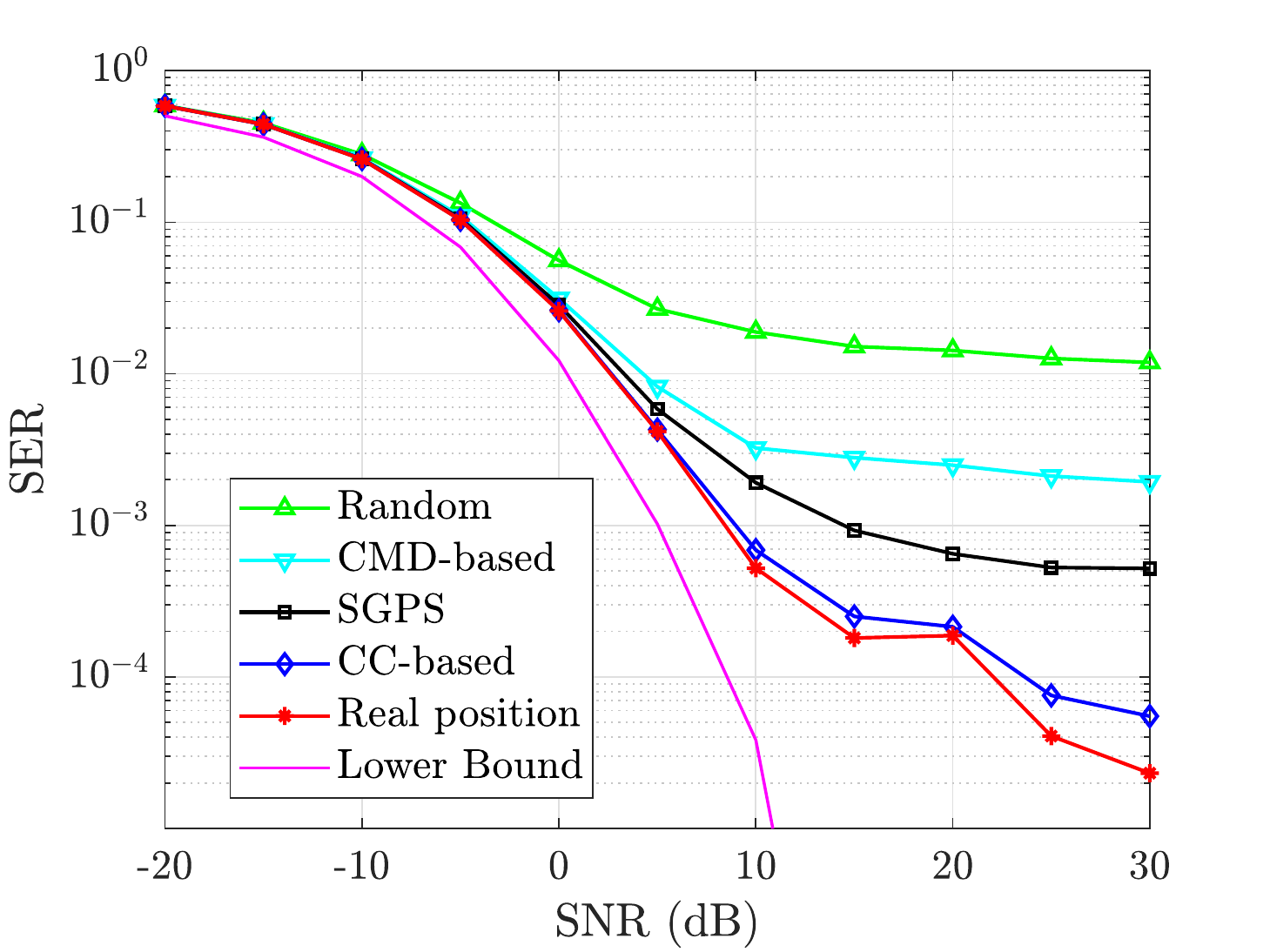}
    \caption{SER as a function of SNR.}
    \label{}
\end{subfigure}
\begin{subfigure}{0.45\textwidth}
    \includegraphics[width=\textwidth,trim=1mm 0mm 10mm 2mm,clip]{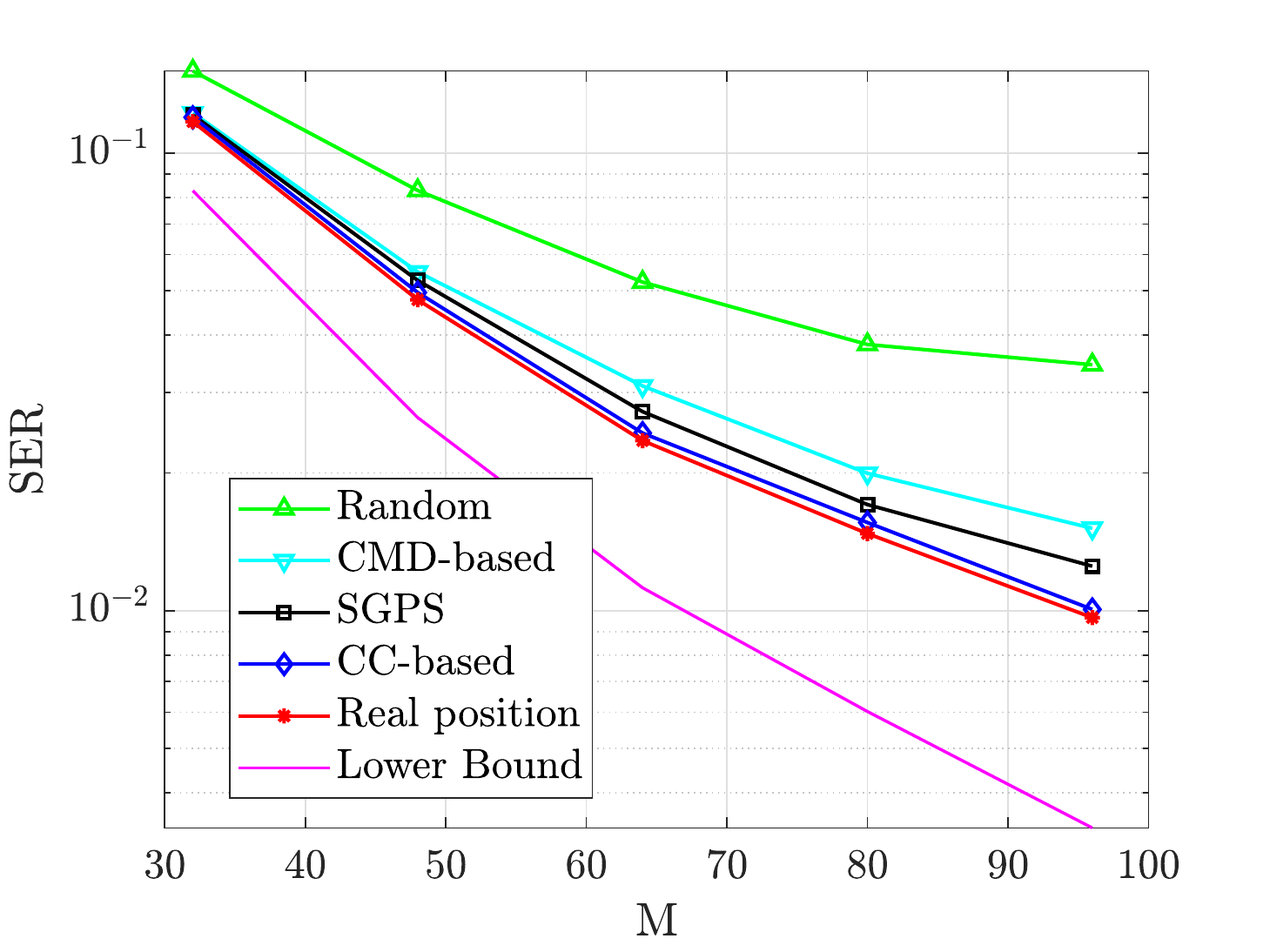}
    \caption{SER as a function of the number of antennas per ULA $M$.}
    \label{}
\end{subfigure}
\caption{Symbol error rate (SER) for a fixed pilot length of $\tau=64$. (a) shows the effect of the SNR and (b) the effect of the number of antennas for a fixed SNR=0~dB.}\vspace{-2mm}
\label{fig:SER}
\end{figure}

\begin{figure}[t]
\centering
\begin{subfigure}{0.45\textwidth}
    \includegraphics[width=\textwidth,trim=1mm 0mm 10mm 2mm,clip]{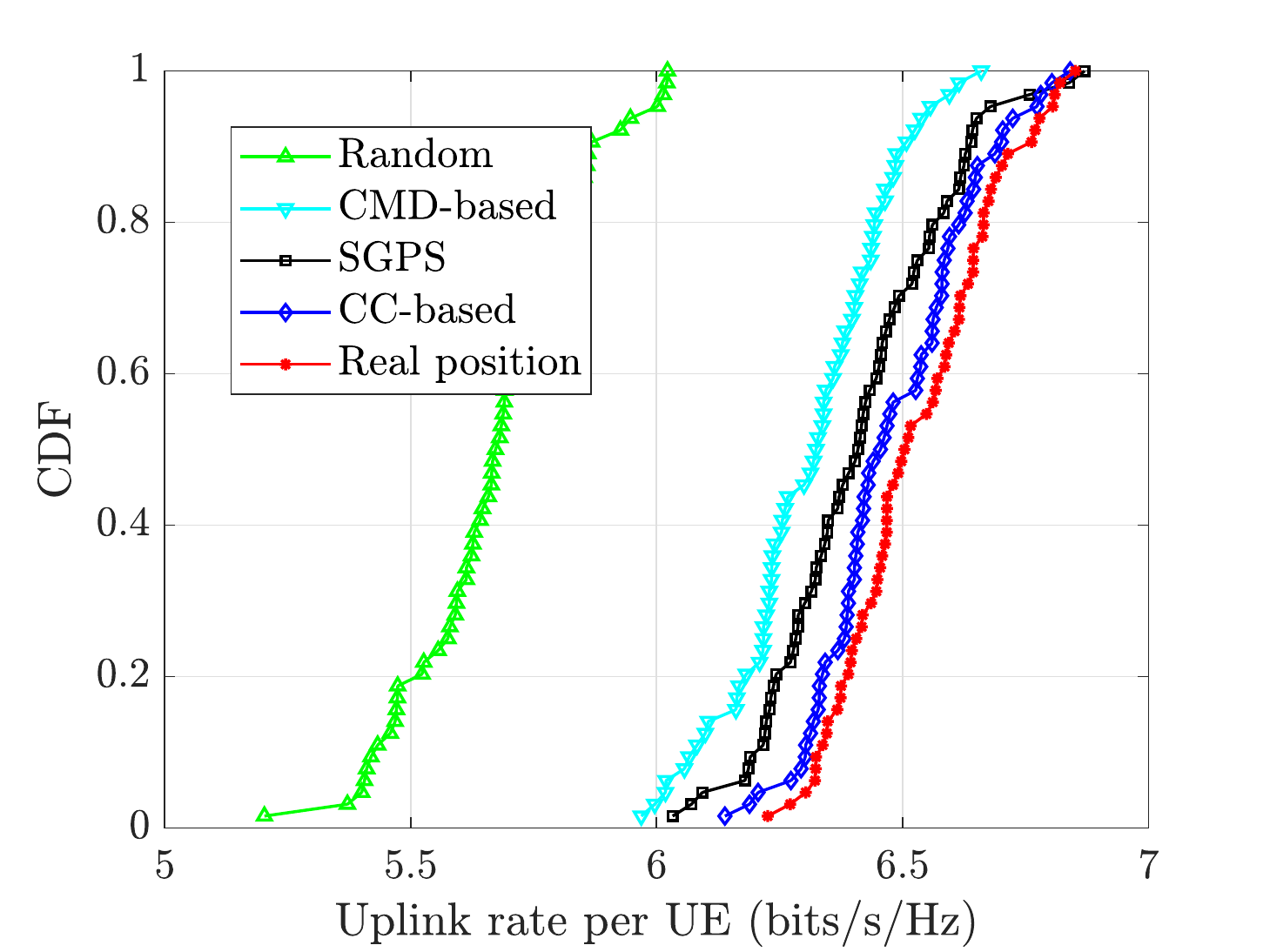}
    \caption{CDF.}
    \label{}
\end{subfigure}
\begin{subfigure}{0.45\textwidth}
    \includegraphics[width=\textwidth,trim=1mm 0mm 10mm 2mm,clip]{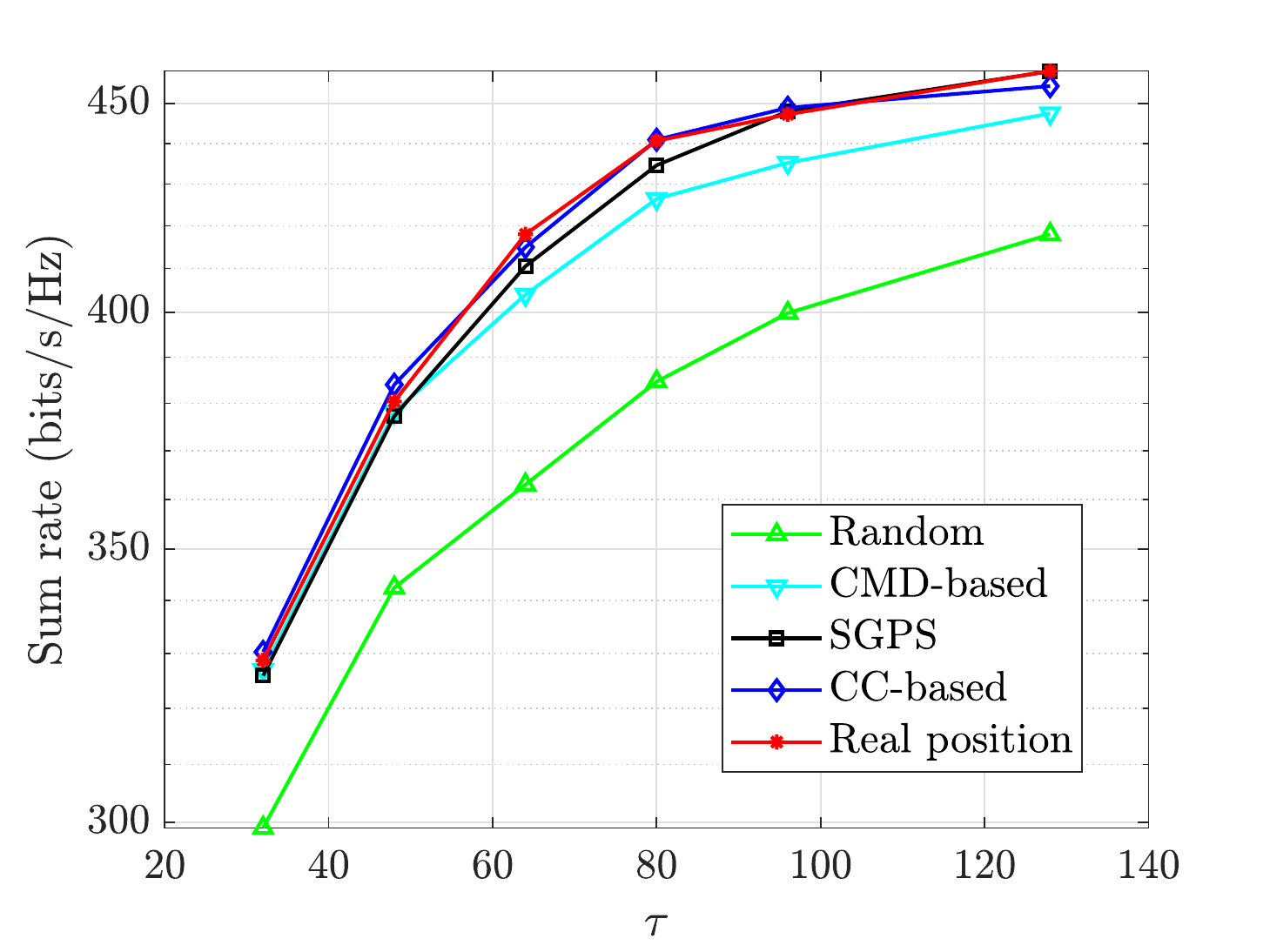}
    \caption{Achievable sum rate.}
    \label{}
\end{subfigure}
\caption{Achievable sum rate at 10~dB SNR. (a) CDF of users' uplink achievable rate for $\tau=64$.(b) Achievable sum rate vs. pilot length.}\vspace{-2mm}
\label{fig:AR}
\end{figure}

We consider three baseline pilot assignment methods: 1) A random pilot assignment scheme; 2) A real position method, which relies on the exact UEs' positions to compute the angular separation between the UEs, and then deploys Algorithm~\ref{alg:NNPA}. Note that this baseline acts as a lower bound to the CC-based method, since it uses the real position instead of the relative position estimated via CC; 3) The statistical greedy pilot scheduling (SGPS) developed in~\cite{You2015} that, similarly to our method, relies on the knowledge of the channel covariance matrices. Differently from~\cite{You2015}, we extract designed features from the channel covariance matrices and then apply Algorithm~\ref{alg:NNPA}; this also obviates the need to store the channel covariance matrices at the BS.

Fig.~\ref{fig:SER} shows how SER of the different pilot assignment strategies changes with the SNR and the number of antennas $M$. A fixed pilot length of $\tau=64$ is considered, which gives a pilot reuse factor of $N/\tau=8$, meaning that each pilot sequence is reused 8 times across all the UEs connected to the BS. For benchmark, we also show the lower bound for SER, which is achieved with perfect CSI knowledge. From Fig.~\ref{fig:SER}(a) we see that the performance of the CC-based pilot allocation approaches the real position method, beating all the other baseline methods. The CMD-based approach, although implementing the same pilot allocation algorithm, shows poor performance as compared to CC-based, which highlights the importance of choosing a good feature to the pilot allocation Algorithm~\ref{subsec:NNPA}. Fig.~\ref{fig:SER}(b) presents the effect of changing the number of ULA elements $M$ for SNR=0~dB. 
By increasing the number of antennas, the performance increases because the UEs' channels become more orthogonal.  

We also evaluate the performance of the proposed features in terms of the uplink achievable rate. The cumulative distribution function (CDF) for the average uplink rate per UE is depicted in Fig.~\ref{fig:AR}(a) for a fixed pilot length of $\tau=64$, at 10~dB SNR. It can be seen that, on average, at least 90~\% of the UEs using the CMD-based pilot allocation strategy have a higher uplink achievable rate than the UE with the highest rate employing the random pilot allocation. On the other hand, this percentage grows to 100~\% when the UEs are employing CC-based or SGPS schemes. 

Fig.~\ref{fig:AR}(b) plots the uplink sum rate against the pilot length. 
One can notice that the CC-based method performs very well, approaching the real position based pilot allocation. However, it slightly degrades its performance for
$\tau>100$ symbols, i.e., when the pilot reuse factor $N/\tau$ decreases below 5. On the other hand, in this regime, the SGPS method performs as well as the real position one. Although the CMD-based method greatly improves the performance as compared to the random pilot allocation, it underperforms when compared to the other methods.  

\section{Conclusions}
This paper investigated the reuse of orthogonal pilot sequences in a single-cell multi-sector network for mMIMO systems. In these systems, the transmission rate performance is interference limited due to pilot contamination. We proposed to apply new features capturing the angular information present in CSI in the algorithm developed in~\cite{Ribeiro2020} to accommodate the multi-sector framework and to improve the average uplink rate for the UEs. The CC-based feature was shown to approach the real position method, outperforming the other baselines in terms of SER and achievable uplink rate. 
The CMD-based feature performs close 
to SGPS and CC-based pilot allocation, whilst having lower complexity. The proposed features showed to fit well into the multi-sector framework by presenting a competitive performance for SER and achievable rate as compared to existing pilot allocation schemes.

\section*{Acknowledgements}

This work has been financially supported in part by the Academy of Finland project ROHM (grants 319484 and 319485). The work of L. Ribeiro, M. Leinonen, and M. Juntti has been financially supported in part by the Academy of Finland 6Genesis Flagship (grant 318927). The work of M. Leinonen has also been financially supported in part by Infotech Oulu and the Academy of Finland (grant 323698).

\bibliographystyle{IEEEtran}
\mybibliography

\end{document}